\documentclass[icol, referee, pdflatex, sn-aps]{sn-jnl}

\usepackage{graphicx}%
\usepackage{multirow}%
\usepackage{amsmath,amssymb,amsfonts}%
\usepackage{amsthm}%
\usepackage{mathrsfs}%
\usepackage[title]{appendix}%
\usepackage{xcolor}%
\usepackage{textcomp}%
\usepackage{manyfoot}%
\usepackage{booktabs}%
\usepackage{algorithm}%
\usepackage{algorithmicx}%
\usepackage{algpseudocode}%
\usepackage{listings}%

\theoremstyle{thmstyleone}%
\theoremstyle{thmstyletwo}%

\theoremstyle{thmstylethree}%

\begin{document}

\title[Article Title]{Deep learning-driven evaluation and prediction of ion-doped NASICON materials for enhanced solid-state battery performance}




\author[1]{\fnm{} \sur{Zirui Zhao}}
\equalcont{These authors contributed equally to this work.}
\author[1]{\fnm{} \sur{Xiaoke Wang}}
\equalcont{These authors contributed equally to this work.}
\author[2]{\fnm{} \sur{Si Wu}}
\author[1]{\fnm{} \sur{Pengfei Zhou}}
\author[1]{\fnm{} \sur{Qian Zhao}}
\author[1]{\fnm{} \sur{Guanping Xu}}
\author[1]{\fnm{} \sur{Kaitong Sun}}
\author*[1]{\sur{Hai-Feng Li}}\email{haifengli@um.edu.mo}
\affil[1]{\orgdiv{Institute of Applied Physics and Materials Engineering}, \orgname{University of Macau}, \orgaddress{\street{Avenida da Universidade}, \city{Taipa}, \postcode{999078}, \state{Macao SAR}, \country{China}}}
\affil[2]{\orgdiv{School of Physical Science and Technology}, \orgname{Ningbo University}, \orgaddress{\city{Ningbo}, \postcode{315211}, \country{China}}}


\abstract{NASICON (Na$_{1+x}$Zr$_2$Si$_x$P$_{3-x}$O$_{12}$) is a well-established solid-state electrolyte, renowned for its high ionic conductivity and excellent chemical stability, rendering it a promising candidate for solid-state batteries. However, the intricate influence of ion doping on their performance has been a central focus of research, with existing studies often lacking comprehensive evaluation methods. This study introduces a deep-learning-based approach to efficiently evaluate ion-doped NASICON materials. We developed a convolutional neural network (CNN) model capable of predicting the performance of various ion-doped NASICON compounds by leveraging extensive datasets from prior experimental investigation. The model demonstrated high accuracy and efficiency in predicting ionic conductivity and electrochemical properties. Key findings include the successful synthesis and validation of three NASICON materials predicted by the model, with experimental results closely matching the model's predictions. \color{black}This research not only enhances the understanding of ion-doping effects in NASICON materials but also establishes a robust framework for material design and practical applications. It bridges the gap between theoretical predictions and experimental validations.}

\keywords{NASICON, solid-state electrolyte, ion doping, deep learning model, electrochemical properties}



\maketitle

\section{Introduction}\label{sec1}

NASICONs (sodium superionic conductors) have emerged as pivotal entities in the realm of solid-state materials due to their remarkable ionic conductivity properties for solid-state batteries \cite{jain2013commentary, fischer2006predicting, hautier2011phosphates, sun2019map, ZHOU2023101248, zhou2023machine}. Belonging to the family of polyanionic compounds, these materials exhibit a distinctive three-dimensional framework structure characterized by interconnected metal oxide polyhedra and dedicated channels facilitating ionic transport \cite{li2023recent}. Seminal work in the 1960s by Hong and Goodenough unveiled the prototype NASICON compound, Na$_3$Zr$_2$(SiO$_4$)$_2$(PO$_4$), highlighting the flexibility in polyanion composition, thereby allowing precise adjustments in the SiO$_4$ to PO$_4$ ratio. This led to the creation of a complete solid-solution series, Na$_{1+x}$Zr$_2$(SiO$_4$)$_x$(PO$_4$)$_{3-x}$ (0 $\leq$ x $\leq$ 3), designed to maintain charge neutrality, achieving an optimal ionic conductivity of 0.2 S$\cdot$cm$^{-1}$ at 300 $^\circ$C, when x = 2. Nevertheless, the intricate interplay between polyanion composition and sodium content in influencing the ionic conductivity necessitates a rigorous decoupling of their discrete contributions \cite{chen2013hollow, goodenough1976fast}.

Furthermore, the NASICON structure has demonstrated its versatility in accommodating an array of polyanion groups such as AsO$_4^{3-}$, SO$_4^{2-}$, and SeO$_4^{2-}$, albeit the exploration of compounds featuring these polyanions has not been as extensive as in the Si system \cite{li2023regulating}. Noteworthy is the NASICON framework's ability to host a diverse range of ions, including divalent, trivalent, tetravalent, and pentavalent metal cations, with relatively high solubilities \cite{jain2016computational}. Employing aliovalent doping strategies has further proved effective in enhancing the ionic conductivity \cite{hautier2011phosphates}.

In this context, this paper aims to delve into the intricate interplay of polyanion composition, sodium content, and their joined influence on the ionic conductivity of NASICON-type materials. Through systematic exploration of these key parameters, we seek to unravel the underlying principles governing the exceptional electrochemical performance exhibited by these compounds, paving the way for tailored advancements in solid-state materials research.

\section{Materials and methods}

\subsection{Deep learning method}

\color{black}In the research of NASICON compounds doped with various elements, we employed a Convolutional Neural Network (CNN) approach as a highly effective methodology \cite{hautier2011phosphates, sun2019map, jain2016computational, Zhang2024, Xu2020, ZHAO2024112982}. CNNs have become central to research in image understanding \cite{chua1998cnn, girshick2015fast}. Their weight-sharing network architecture closely resembles biological neural networks, reducing the complexity of the model \cite{alzubaidi2021review, bhatt2021cnn}. CNNs offer significant advantages over traditional neural networks, primarily through their distinct layer structure, where each layer performs a specialized function \cite{Ren2015faster}. CNNs are widely applied in materials science, particularly for predicting and optimizing compound properties. By leveraging deep learning techniques, this method autonomously learns and extracts features inherent to compounds, providing a powerful tool for analyzing and understanding the impact of various element dopants on NASICON compounds' properties \cite{davies2016computational, soundharrajan2023exploring}.\color{black}

\color{black} Moreover, CNNs offer a more detailed analysis of element doping effects, including precise lattice positions and interaction mechanisms, due to their capability to automatically capture complex features of compound lattice structures and element distributions, eliminating the need for extensive manual feature engineering \cite{mandal2023machine, zhang2023revolutionizing, zhou2023machine, nakano2021molecular, fukuda2022bayesian}.\color{black}

Furthermore, the high parallelism and adaptability of CNNs allow for the processing of extensive datasets, thus making high-throughput materials screening and design feasible \cite{iwamizu2023search, takeda2022process, zhang2024mining, wang2020lithium}.

By employing CNNs, we systematically investigate the influence of different elements' doping on the properties of NASICON compounds. This approach provides profound insights into materials design and development, equipping us with a robust tool for exploring potential performance enhancements and expanding applications of NASICON-type compounds.

\subsection{Synthesis of solid NASICON electrolytes}

The synthesis of our materials involved the utilization of a solid-state method \cite{li2008synthesis, Li2007-2}. Raw materials, namely Na$_3$PO$_4$, ZrO$_2$, and CuO$_2$, Fe$_2$O$_3$, Nd$_2$O$_3$ were employed in the process. An additional 15wt\% of Na$_3$PO$_4$ was incorporated. The milling phase was carried out for one hour using a Vibratory Micro Mill. Subsequently, calcination was performed at a temperature of 1100 $^{\circ}$C for a duration of 12 hours, employing a heating rate of 3 $^{\circ}$C/min. For pellet formation, a pressure of 800 MPa was applied, with 0.6 gram of material per pellet. The sintering process was conducted at 1200 $^{\circ}$C for 12 hours, employing the same heating and cooling rates. The incorporation of excess Na$_3$PO$_4$ and the synthesis parameters were adopted from our previous studies \cite{ZHOU2023101248, zhou2023machine}.

\subsection{Electrochemical property measurements}

In the realm of electrochemical measurement, electrochemical impedance spectroscopy (EIS) was the chosen technique. This was executed using an electrochemical workstation, specifically the ModuLab XM. An amplitude of 5 mV was applied, covering a frequency range from 1 MHz down to 1 Hz. Blocking electrodes were introduced, comprising a thin layer of Ag sputtered on both sides of the pellets with a vacuum vapor deposition. We conducted electrochemical impedance spectroscopy at room temperature following the procedure outlined in \cite{ZHOU2023101248}. The total conductivity was computed using the formula:
\begin{eqnarray}
\sigma = \frac{1}{R}\frac{L}{S},
\label{total}
\end{eqnarray}
where \emph{R} represents the total resistance, \emph{L} denotes the pellet thickness, and \emph{S} signifies the area of the blocking electrode. The fitted impedance was calculated by the equation~\ref{total}.

\section{Results and discussion}

\subsection{Data collection}

\begin{figure*} [!t]
\centering \includegraphics[width=0.82\textwidth]{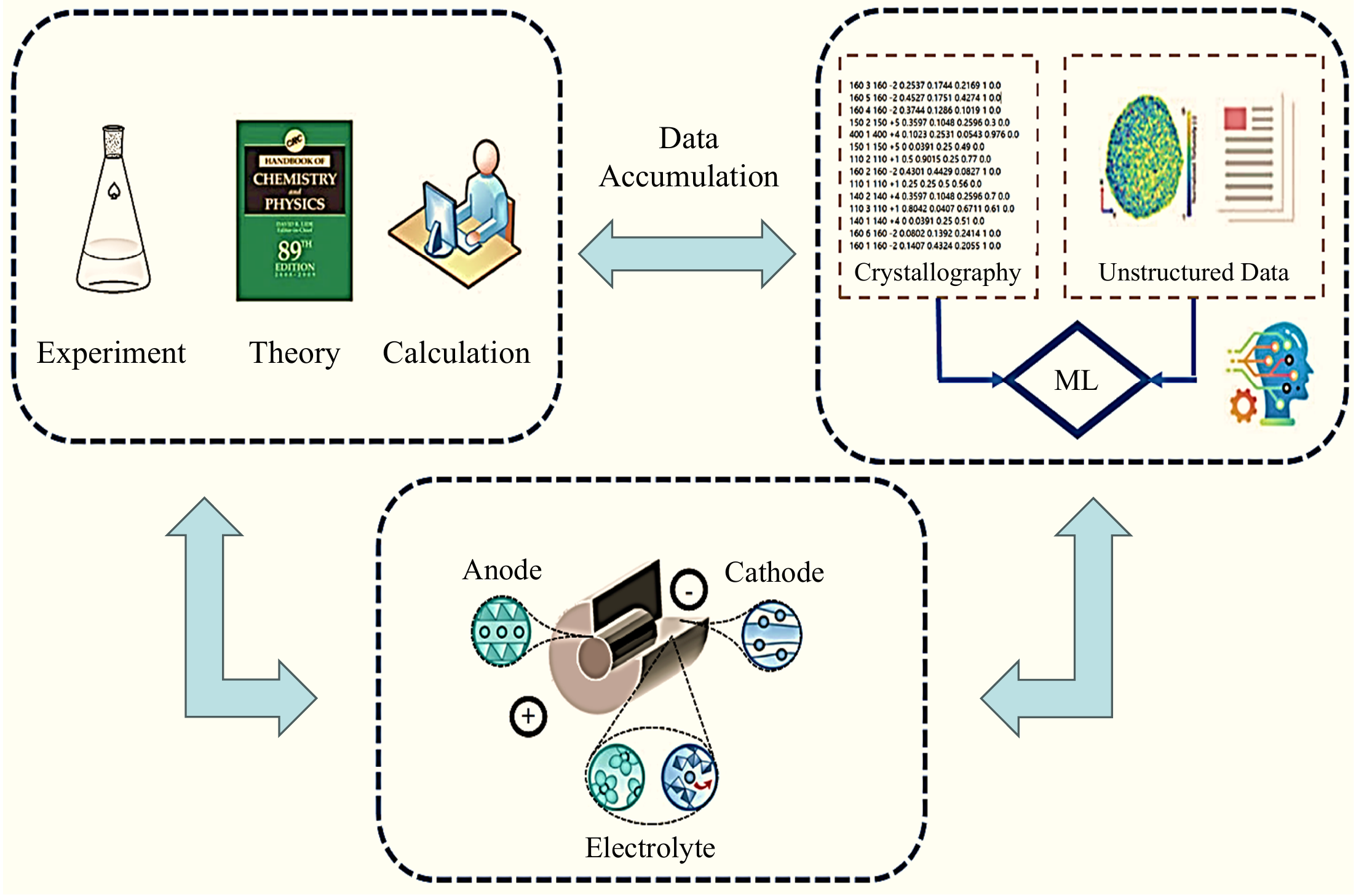}
\caption{Data-driven workflow for predicting NASICON material components and properties. Previous research findings were collected and structured within our artificial neural network (ANN) to facilitate training and assess its robustness in both the training and test sets.}
\label{workflow}
\end{figure*}

The workflow for NASICON electrode prediction, as depicted in Fig.~\ref{workflow}, was meticulously executed. Approximately 6,000 CIF files of diverse NASICON-type materials dating back to 1960 were systematically collected and crawled from websites using natural language processing techniques \cite{MSIEureka2015-1, MSIEureka2012-2, MSIEureka2011-3, Villars2023-4, Villars2023-5, Villars2023-6, Villars2023-7, MSIEureka2016-8, Villars2023-9, MSIEureka2011-10, Villars2023-11, Villars2023-12, Villars2023-13, jain2013commentary}. These files were meticulously correlated with their respective electrochemical properties, culminating in the establishment of a dedicated database, which serves as a foundational resource for our subsequent investigations. Subsequently, CNNs were leveraged for deep learning on this database, with the objective of enabling the direct prediction of diverse outcomes in NASICON compounds with varying element doping, solely by inputting CIF files.

The CIF files used in this study were obtained from reputable open-access websites, scientific databases, and peer-reviewed publications where data sharing is explicitly permitted. These files were sourced from well-known databases such as the Crystallography Open Database (COD) and the Materials Project, which are recognized for their accuracy and reliability. To ensure the integrity of the data, the CIF files were cross-verified against multiple sources. All data collection methods strictly adhered to legal and ethical guidelines, ensuring that the datasets used are freely available for research purposes. Therefore, the CIF files utilized in this study are reliable and legally obtained.

\begin{figure*} [!t]
\centering \includegraphics[width=0.82\textwidth]{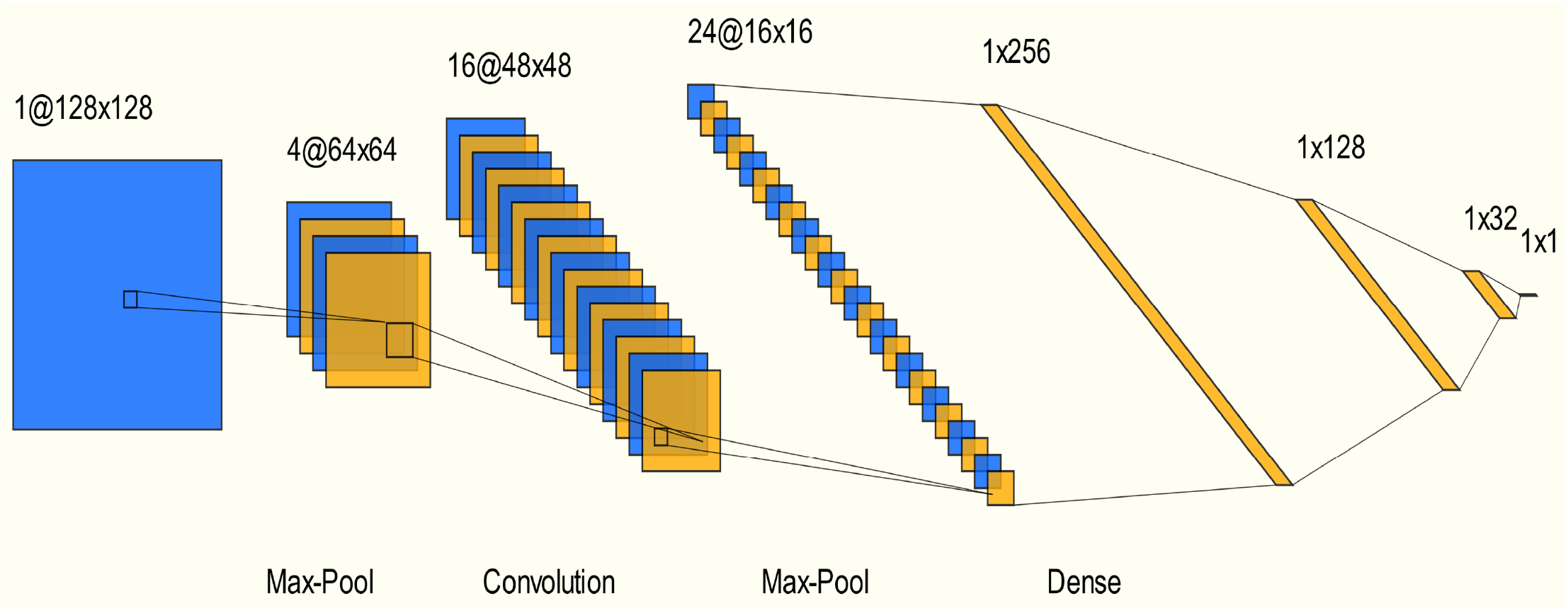}
\caption{Artificial convolutional neural network consisting of 13 layers, constructed based on an adjusted sample size of collected data. The network comprises 3 convolutional layers, 3 pooling layers, and 7 fully-connected layers to generate the final data output.
}
\label{layers}
\end{figure*}

Figure~\ref{layers} illustrates the artificial convolutional neural network with a total depth of 13 layers. (1) The input layer: In machine learning, data preprocessing is essential to optimize model performance, mitigate data irregularities, and ensure convergence speed. Data preprocessing primarily involves descriptor screening, with principal component analysis utilized to reduce data dimensionality, resulting in the identification of five independent descriptors \cite{mackiewicz1993principal, yang2004two}. Descriptor weights were determined through correlation analysis, enhancing prediction efficiency. (2) Convolutional layer: This pivotal layer excels at extracting features from two-dimensional structured data. Comprising multiple convolutional units, parameterized through backpropagation, it executes diverse operations. Convolution kernels, usually square and linear, manipulate input image data and the convolution kernel matrix to traverse pixels and calculate weighted sums. The convolution process, involving bias addition, generates the output feature image \cite{roska1993cnn, girshick2015fast}. (3) Activation layer: Nonlinear correlation operations are essential due to the inherent limitations of linear models. Activation functions play a pivotal role in facilitating this nonlinearity, elevating the expressiveness of the network model. Commonly used activation functions include ReLU, Sigmoid, and Tanh. For this study, the ReLU function was chosen for its computational convenience \cite{girshick2015fast}. (4) Pooling layer: These layers streamline network structures by reducing feature image dimensionality, thus mitigating overfitting and enhancing generalization. Pooling compresses features while retaining vital information, contributing to local linear transformation invariance. Typical pooling methods include Average and Max pooling \cite{chua1998cnn, girshick2015fast}. (5) Fully-connected layer: This layer primarily addresses classification tasks within neural network models. It applies weights to previous layer features, mapping distributed feature representations to the sample labeling space \cite{chua1998cnn, girshick2015fast}.

\subsection{Model setup}

For this study, the PyTorch framework was employed in the deep learning system to process various CIF files from a database. A CNN model inspired by the classical VGG16 architecture, resulting in a 13-layer network, was constructed (see Fig.~\ref{layers}). This model was evaluated and scored in the context of NASICON. Relevant model parameters are presented in Table~\ref{parameters}, where epoch denotes the number of training iterations; batch size specifies the number of files used for training; learning rate governs information accumulation in the neural network over time, with lower values causing slower gradient descent and impacting convergence behavior; and cross-entropy loss serves as the loss function to minimize differences between predicted and true values, 
\begin{equation}
\textrm{Loss} = -\sum_{i}^{N} p(x_i) \log q(x_i),
\end{equation}
where $p(x_i)$ represents the true distribution, $q(x_i)$ denotes the predicted distribution, and $x_i$ is the random variable. The model has been rigorously trained and meticulously validated to ensure its accuracy, reliably forecasting the properties of NASICON compounds. 

\subsection{Database integration and predictive modeling}

Our approach commenced with the meticulous assembly of a comprehensive database containing approximately 6,000 CIF files of diverse NASICON compounds. This exhaustive dataset was curated from various sources, encompassing decades of research in the realm of solid-state materials. The database incorporates an extensive array of NASICON compositions, providing a rich foundation for predictive modeling. To enhance the model's predictive capabilities, the CIF files were associated with their corresponding electrochemical properties, establishing a robust training dataset. Each entry in the database encapsulates critical information about the chemical composition, crystal structure, and ionic conductivity of NASICON compounds. This holistic integration reflects the diversity of NASICON materials and ensures a nuanced understanding of the relationships between composition and performance. The predictive modeling phase involved the utilization of CNNs, a cutting-edge deep learning architecture well-suited for analyzing structured data like crystallographic information. The CNN was trained on the integrated database, learning intricate patterns and correlations between the structural features of NASICON compounds and their electrochemical behavior.

To ensure robust model performance and prevent overfitting, we employed a 5-fold cross-validation technique. The dataset \( D \) was randomly partitioned into five equally sized subsets \( D_1, D_2, D_3, D_4, D_5 \). In each iteration of the cross-validation, four subsets were combined to form the training set \( D_{\text{train}} \) (e.g., \( D_{\text{train}} = D_1 \cup D_2 \cup D_3 \cup D_4 \)) while the remaining subset was used as the test set \( D_{\text{test}} \) (e.g., \( D_{\text{test}} = D_5 \)). This process was repeated five times, with each subset serving as the test set exactly once. Using 80\% of the data for training (\( D_{\text{train}} \)) and 20\% for testing (\( D_{\text{test}} \)) ensures that the model is trained on a substantial portion of the data, while the evaluation on \( D_{\text{test}} \) provides a reliable measure of the model’s generalization ability. The mean performance metric across all five folds was used to assess the model’s overall effectiveness and robustness.

\begin{figure*} [!t]
\centering
\includegraphics[width=0.82\linewidth\hspace{-5mm}]{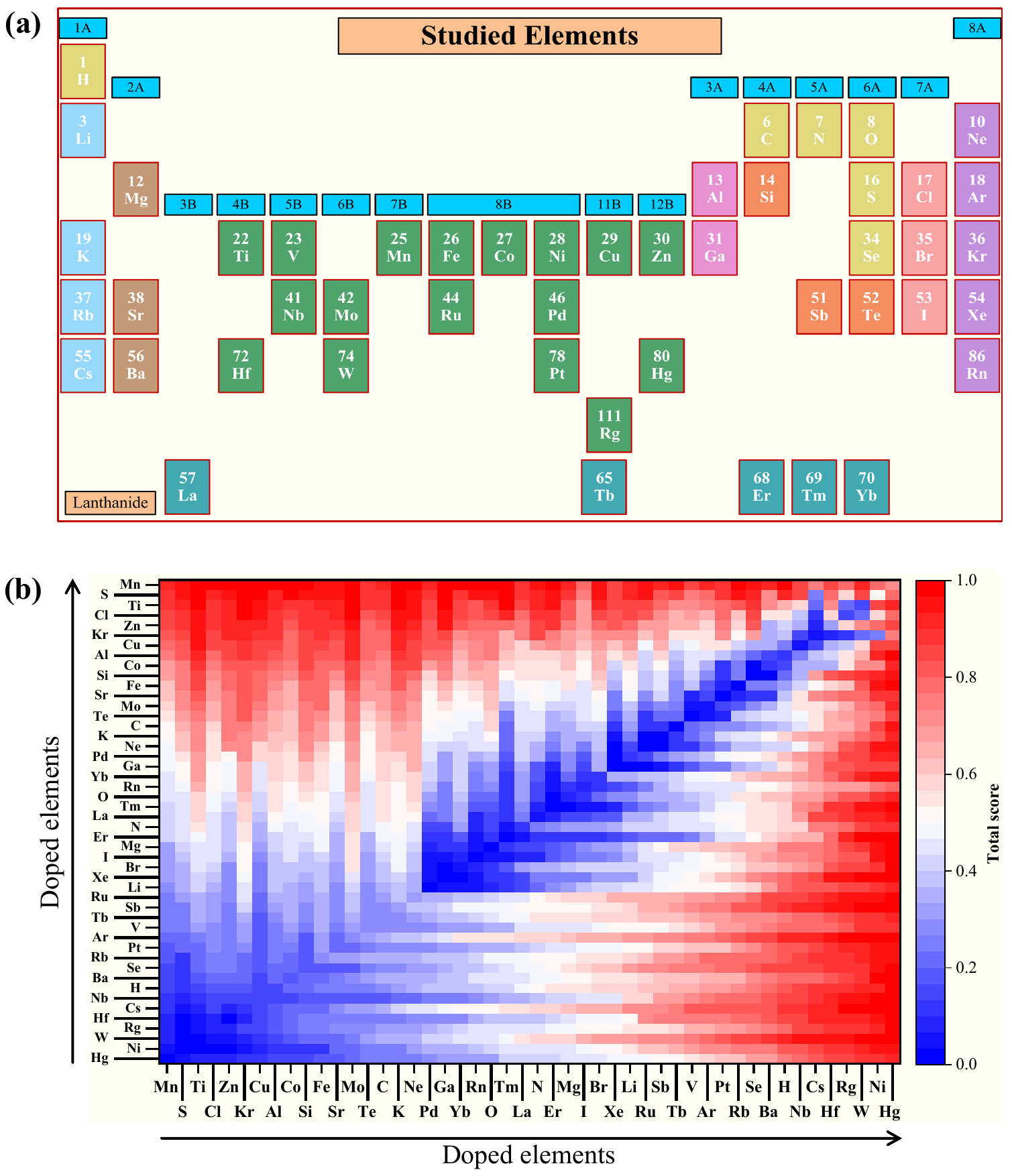}
\caption{\color{black}\textbf{a} The positions of forty-nine elements on the periodic table of chemistry. 
\textbf{b} Scores of forty-nine different elements in the double-doping model, as output by our trained model after normalization, considering descriptors weighted as follows: 85$\%$ for ionic conductivity, 5$\%$ for the electrochemical stability window, 5$\%$ for thermal stability, and 5$\%$ for the chemical diffusion coefficient.}
\label{scores}
\end{figure*}

Our model's proficiency underwent systematic validation through cross-validation techniques, ensuring its robustness in predicting NASICON compositions with superior electrochemical properties. Through this partitioning, we attained an accuracy of 0.91 on the training set and 0.85 on the test set. In this study, forty-nine distinct elements from the periodic table of chemistry (Fig.~\ref{scores}a) were selected to investigate their doping effect on NASICON performance. The successful integration of this comprehensive database with advanced predictive modeling represents a paradigm shift in the expedited discovery of high-performance NASICON materials (Fig.~\ref{scores}b).

\subsection{Predicted and tested compounds}

Among the 49 compounds analyzed, 17 exhibited superior performance compared to the originally reported ion-doped elements. It is important to note that these 17 compounds were not merely those that matched the initial predictions; rather, they demonstrated enhanced performance metrics. Furthermore, we proposed new doping ratios for these compounds to further optimize their performance.

\begin{figure*} [!t]
\centering \includegraphics[width=0.82\textwidth]{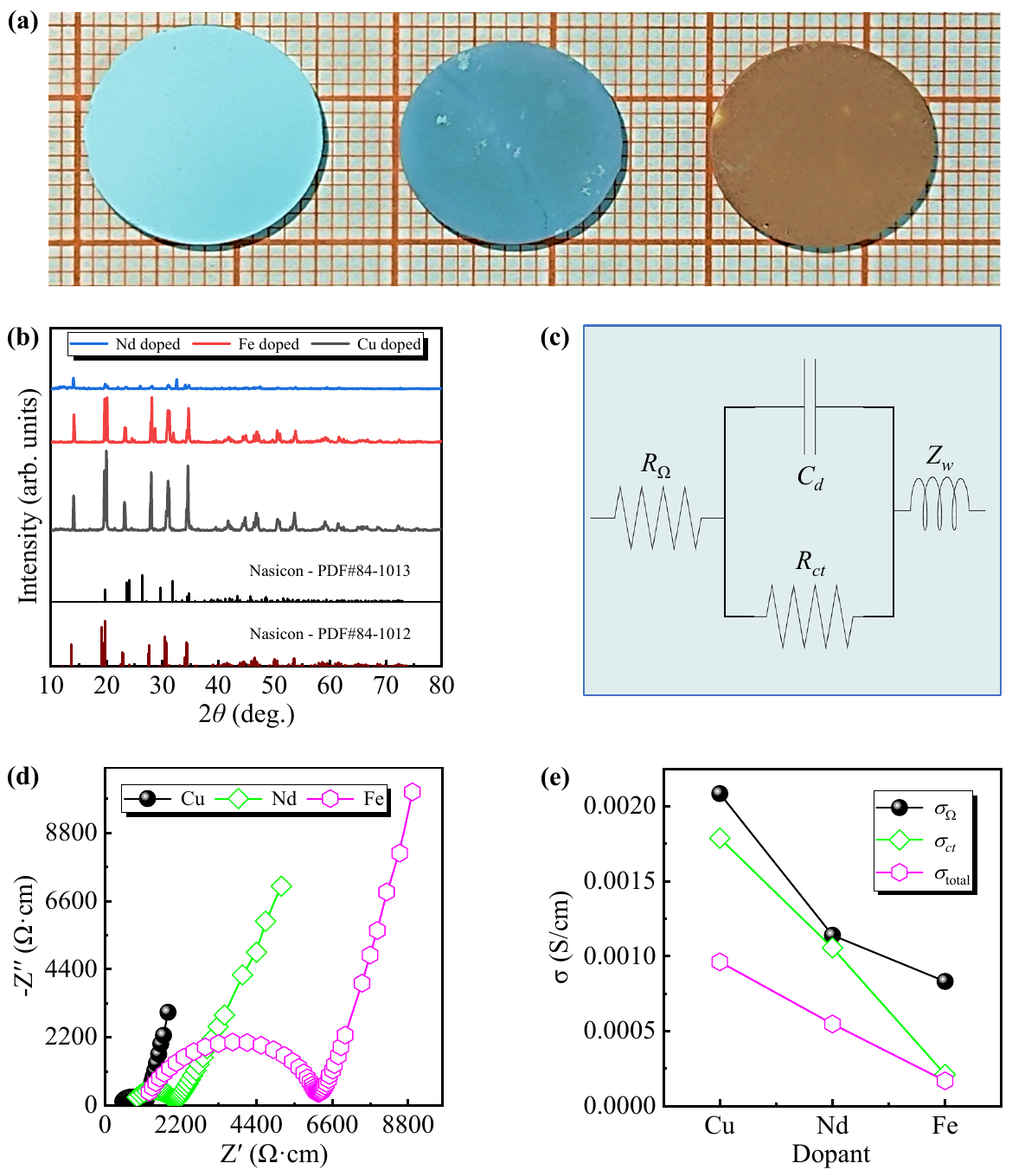}
\label{samples}
\caption{
\textbf{a} Synthesised solid NASICON electrolytes, arranged from left to right: Na$_{3.36}$Zr$_{1.92}$Cu$_{0.08}$Si$_{2.2}$P$_{0.1}$O$_{12}$ (light azure blue), Na$_{3.1}$Zr$_{1.9}$Nd$_{0.1}$Si$_{2}$PO$_{12}$ (dark azure blue), and Na$_{3.2}$Zr$_{1.8}$Fe$_{0.2}$Si$_{2}$PO$_{12}$ (dark brown). 
\textbf{b} Corresponding X-ray powder diffraction patterns of the three NASICON electrolytes.
\textbf{c} The equivalent circuit diagram utilized in the EIS measurement, where $R_{\Omega}$ denotes the internal resistance, $C_{d}$ represents the double-decker capacitance, $R_{ct}$ signifies the charge transfer resistance, and $Z_{w}$ denotes the Warburg impedance.
\textbf{d} Nyquist plots illustrating Cu-doped (black balls), Nd-doped (green squares), and Fe-doped (pink hexagons) NASICON electrodes. Based on the measured geometrical factors, we calculated the room-temperature impedance as resistivity. 
\textbf{e} Calculated conductivities of $\sigma_{\Omega}$, $\sigma_{ct}$, and $\sigma_{\textrm{total}}$. 
The radius of all synthesized solid NASICON electrolytes was 0.8 cm, while the thicknesses of Cu-doped, Fe-doped, and Nd-doped solid NASICON electrolytes were 0.127 cm, 0.112 cm, and 0.113 cm, respectively.}
\label{test}
\end{figure*}

Therefore, we believe that our predictive model successfully identified and recommended 16 NASICON compounds with outstanding properties (see Table~\ref{predication}). Among them, the Co-doped NASICON compound exhibits the best predicated performance. Specifically, the Na$_{3.3}$Zr$_{1.95}$Co$_{0.05}$Si$_{2.2}$P$_{0.8}$O$_{12}$ sample synthesized demonstrated the highest ionic conductivity of $\sim 2.63 \times 10^{-3}$ S/cm \cite{zhou2023machine}. The NASICON materials doped with Cu, Nd, and Fe (as depicted in Fig.~\ref{test}a) underwent X-ray powder diffraction and EIS characterizations. X-ray powder diffraction studies reveal their crystalline nature as a single phase (Fig.~\ref{test}b).

To ensure the validity of our predicted outcomes, an EIS test was conducted, and the corresponding equivalent circuit diagram for the specific measurement is depicted in Fig.~\ref{test}c. The overall impedance of the system can be decomposed into three electrical components: internal resistance ($R_{\Omega}$), double-decker capacitance ($C_{d}$), and Faraday impedance ($Z_{f}$), where the Faraday process further comprises charge transfer and mass transfer, abstracted as charge transfer resistance ($R_{ct}$) and Warburg impedance ($Z_{w}$), respectively. The impedance of the entire system, as measured by EIS, can be expressed as a complex function \( Z(\omega) \), with \( \omega \) representing angular frequency. This impedance function typically comprises real and imaginary components, denoting resistance and reactance respectively:
\begin{equation}
Z(\omega) = Z_{Re} + jZ_{Im},
\end{equation}
where $Z_{Re}$ denotes the real part and $Z_{Im}$ represents the imaginary part of the impedance, with \( j \) as the imaginary unit. Based on the provided equivalent circuit diagram (Fig.~\ref{test}c), we can describe the impedance of the entire measurement system using Ohm's law, yielding the following expressions for the real and imaginary parts:
\begin{equation}
Z_{Re} = R_{\Omega} + \frac{R_{ct} + \sigma \omega^{-1/2}}{(C_{dl}\sigma \omega^{1/2} + 1)^{2} + \omega^{2}C_{dl}^{2}(R_{ct} + \sigma \omega^{-1/2})^{2}},
\label{Re}
\end{equation}
and
\begin{equation}
Z_{Im} = \frac{\omega C_{d}(R_{ct} + \sigma \omega)^{-\frac{1}{2}} + \sigma \omega (i \omega)^{\frac{1}{2}}C_{d}\sigma + 1}{((C_{d}\sigma \omega)^{\frac{1}{2}} + 1)^2 + \omega^2 C_{d}^2(R_{ct} + \sigma \omega)^{-1}}.
\label{Im}
\end{equation}
It is worth noting that equations~\ref{Re} and \ref{Im} are computationally complex. Hence, for practical purpose, we consider two limiting cases: when frequency tends to 0, the equation simplifies to:
\begin{equation}
Z_{Im} = Z_{Re} - R_{\Omega} - R_{ct} + 2\sigma^2C_{d}.
\label{Sim-1}
\end{equation}
From the simplified equation~\ref{Sim-1}, it is apparent that it should exhibit a linear relationship with a slope of one and intersect the real axis at ($R_{\Omega} + R_{ct}$). Similarly, as frequency tends to infinity, the equation can be simplified to:
\begin{equation}
\left( Z_{Re} - R_{\Omega} - \frac{R_{ct}}{2} \right)^2 + Z_{Im}^2 = \left( \frac{R_{ct}}{2} \right)^2.
\label{Sim-2}
\end{equation}
Equation~\ref{Sim-2} describes a semicircle, corresponding to the arc observed in the first half of the measurements, as depicted in Fig.~\ref{test}d, with its centre and radius defined as ($R_{\Omega} + R_{ct}/2$) and $R_{ct}/2$, respectively.

Upon fitting the EIS test data to determine the resistance values $R_{\Omega}$ and $R_{ct}$ within the entire system, the total conductivity of each sample can be individually calculated using equation~\ref{total}. For practical considerations, we categorize the frequency range as follows: high frequency range $10^{3} \sim 10^{6}$ Hz, and low to medium frequency range as $10^{-3} \sim 10^{3}$ Hz. Following the completion of the fitting process, the values of $R_{\Omega}$, $R_{ct}$, and the final total conductivity ($\sigma_{\textrm{total}}$) of the synthesized materials are presented in Table~\ref{Rvalue}. 

While the three doped NASICON compounds synthesized in this study have been reported in previous literature, our research introduces 17 new doping ratios and successfully synthesizes three of these, all demonstrating enhanced performance. This study provides specific compositions that offer practical guidance for material optimization, rather than merely predicting suitable elements. Compared to previous studies \cite{wang2023design}, our work offers concrete and actionable insights, bridging the gap between theoretical predictions and practical applications.

As illustrated in Fig.~\ref{test}e and Table~\ref{Rvalue}, for Cu-doped NASICON, our material (Na$_{3.36}$Zr$_{1.92}$Cu$_{0.08}$Si$_{2.2}$P$_{0.8}$O$_{12}$) exhibited an enhanced ionic conductivity of $9.62 \times 10^{-4}$ S/cm, surpassing the reported value of $3.38 \times 10^{-4}$ S/cm \cite{wang2023enhanced}. Similarly, our Nd-doped NASICON (Na$_{3.1}$Zr$_{1.9}$Nd$_{0.1}$Si$_{2}$PO$_{12}$) exhibited an enhanced ionic conductivity of $5.48 \times 10^{-4}$ S/cm, outperforming the reported value of $4.21 \times 10^{-7}$ S/cm \cite{barre2008double}. Furthermore, our Fe-doped NASICON (Na$_{3.2}$Zr$_{1.8}$Fe$_{0.2}$Si$_{2}$PO$_{12}$) demonstrated higher ionic conductivity, measuring $1.68 \times 10^{-4}$ S/cm, compared to the previously reported value of $1 \times 10^{-7}$ S/cm \cite{chung2018rhombohedral}. 

\color{black}These results affirm the effectiveness of our database integration and predictive modeling approach in acquiring NASICON compounds with noteworthy ionic conductivities. The observed increase in ionic conductivity in the synthesized ion-doped NASICON materials is particularly significant, as it indicates enhanced charge carrier mobility within the lattice structure. This improvement in ionic conductivity is intrinsically linked to greater efficiency in ion transport, which is crucial for the performance of solid-state electrolytes in energy storage and conversion applications. The enhancement suggests that the doped ions effectively optimize the ion migration pathways, thereby lowering the energy barriers associated with ion transport. Furthermore, the superior electrochemical performance exhibited by our synthesized materials, compared to existing literature, further validates the reliability and robustness of our model in accurately predicting and optimizing material properties. These findings not only underscore the promising potential of the synthesized NASICON compounds for enhanced ionic conductivity but also emphasize their broader implications for advanced technological applications. The significant improvements in ionic transport efficiency suggest that these materials could play a critical role in developing next-generation solid-state electrolytes, particularly in energy storage and conversion systems. By demonstrating both high performance and reliability, these NASICON compounds emerge as strong candidates for integration into cutting-edge devices, further validating the approach used in their design and synthesis.\color{black}

\section{Conclusions}

This study presents a novel deep learning approach to evaluate and predict the performance of ion-doped NASICON materials, which are crucial for advancing solid-state battery technology. By developing a Convolutional Neural Network (CNN) model, we were able to leverage extensive datasets from prior experimental studies to accurately predict the ionic conductivity and electrochemical properties of various ion-doped NASICON compounds. The research methods included automated retrieval and analysis of CIF files using CNNs and natural language processing, thereby creating a comprehensive database linking CIF files to electrochemical properties.

Key findings of the study include the successful synthesis and experimental validation of three high-performing NASICON materials, which were accurately predicted by the CNN model. The experimental results closely matched the model's predictions, demonstrating the model's high accuracy and reliability. This approach not only enhances the understanding of the effects of ion doping in NASICON materials but also provides a robust framework for material design and practical applications.

In conclusion, this manuscript significantly contributes to the field of solid-state electrolytes by bridging the gap between theoretical predictions and experimental validations. The deep learning model developed in this study offers a powerful tool for predicting and optimizing the performance of ion-doped NASICON materials, paving the way for the development of more efficient and reliable solid-state batteries.

\clearpage

\backmatter

\section*{Acknowledgments}

This work was supported by the Science and Technology Development Fund, Macao SAR (File No. 0090{/}2021{/}A2) and the Guangdong{-}Hong Kong{-}Macao Joint Laboratory for Neutron Scattering Science and Technology (Grant No. 2019B121205003).

\section*{Authors’ contributions}

Zirui Zhao and Xiaoke Wang contributed equally. 
Zirui Zhao: Conceptualization, data curation, formal analysis, investigation, visualization, writing-original draft. 
Xiaoke Wang: Conceptualization, data curation, formal analysis, investigation, visualization, writing-original draft. 
Si Wu: Formal analysis, investigation, methodology, visualization. 
Pengfei Zhou: Formal analysis, investigation, methodology, visualization. 
Qian Zhao: Formal analysis, investigation, methodology, visualization. 
Guanping Xu: Formal analysis, investigation, methodology, visualization. 
Kaitong Sun: Formal analysis, investigation, methodology, visualization. 
Hai-Feng Li: Conceptualization, funding acquisition, methodology, project administration, supervision, visualization, writing-review \& editing.

\section*{Availability of data and materials}

All data generated or analyzed during this study are included in this published article.

\section*{Declarations}

\textbf{Ethics approval and consent to participate}

The authors declare they have upheld the integrity of the scientific record.

\textbf{Consent for publication}

The authors give their consent for publication of this article.

\textbf{Competing interests}

The authors declare that they have no competing interests.

\textbf{Non-financial interests}

Hai-Feng Li serves as an editor for the AAPPS Bulletin

\textbf{Financial interests}

The authors declare they have no financial interests.

\clearpage

\bibliographystyle{sn-aps}
\bibliography{Zirui-1.bib}

\clearpage

\clearpage

\clearpage

\clearpage

\clearpage

\begin{table}[!t]
\centering
\caption{Neural network model parameters.}
\label{parameters}
\begin{tabular}{ll}
\hline
Parameter                           & Value                            \\
\hline
Epoch                               & 100                              \\
Batch size                          & 200                              \\
Learning rate                       & 0.0001                           \\
Loss function                       & Cross-entropy loss               \\
Average single epoch time           & 135.3 s                          \\
\hline
\end{tabular}
\end{table}

\clearpage

\begin{table}[!t]
\small
\centering
\caption{Predicated solid-state NASICON electrolytes with potential outstanding properties.}
\label{predication}
\setlength{\tabcolsep}{8.8mm}{}
\renewcommand{\arraystretch}{1.1}
\begin{tabular}{llll}
\hline
\hline
Chemical composition                                                    & Dopant ion                        & Total Score                \\
\hline
Na$_{3.4}$Zr$_{1.8}$Co$_{0.2}$Si$_{2}$PO$_{12}$                         & Co$^{2+}$                         & 0.89                       \\
Na$_{3.36}$Zr$_{1.92}$Cu$_{0.08}$Si$_{2.2}$P$_{0.8}$O$_{12}$            & Cu$^{2+}$                         & 0.87                       \\
Na$_{3.4}$Zr$_{1.8}$Ca$_{0.2}$Si$_{2}$PO$_{12}$                         & Ca$^{2+}$                         & 0.87                       \\
Na$_{3.1}$Zr$_{1.95}$Mg$_{0.05}$Si$_{2}$PO$_{12}$                       & Mg$^{2+}$                         & 0.86                       \\
Na$_{3.46}$Zr$_{1.9}$Mg$_{0.1}$Si$_{2.23}$P$_{0.77}$O$_{12}$            & Mg$^{2+}$                         & 0.86                       \\
Na$_{3.1}$Zr$_{1.9}$Nd$_{0.1}$Si$_{2}$PO$_{12}$                         & Nd$^{3+}$                         & 0.85                       \\
Na$_{3.2}$Zr$_{1.8}$Fe$_{0.2}$Si$_{2}$PO$_{12}$                         & Fe$^{3+}$                         & 0.85                       \\
Na$_{3.4}$Zr$_{1.8}$Ni$_{0.2}$Si$_{2}$PO$_{12}$                         & Ni$^{2+}$                         & 0.84                       \\
Na$_{3.2}$Zr$_{1.8}$Al$_{0.2}$Si$_{2}$PO$_{12}$                         & Al$^{3+}$                         & 0.84                       \\
Na$_{3.4}$Zr$_{1.85}$Zn$_{0.15}$Si$_{2}$PO$_{12}$                       & Zn$^{2+}$                         & 0.83                       \\
Na$_{3.1}$Zr$_{1.9}$La$_{0.1}$Si$_{2}$PO$_{12}$                         & La$^{3+}$                         & 0.83                       \\
Na$_{3.4}$Sc$_{2}$Si$_{0.4}$P$_{2.6}$O$_{12}$                           & Sc$^{3+}$                         & 0.82                       \\
Na$_{3.1}$Zr$_{1.9}$Y$_{0.1}$Si$_{2}$PO$_{12}$                          & Y$^{3+}$                          & 0.82                       \\
Na$_{3}$Zr$_{1.88}$Y$_{0.12}$Si$_{2}$PO$_{12}$                          & Y$^{3+}$                          & 0.82                       \\
Na$_{3}$Zr$_{1.9}$Ce$_{0.1}$Si$_{2}$PO$_{12}$                           & Ce$^{4+}$                         & 0.81                       \\
Na$_{3}$Zr$_{2}$Ge$_{0.15}$Si$_{1.85}$PO$_{12}$                         & Ge$^{4+}$                         & 0.80                       \\
Na$_{3.4}$Zr$_{2}$Si$_{2.4}$P$_{0.6}$O$_{12}$                           & Si$^{4+}$                         & 0.80                       \\
\hline
\hline
\end{tabular}
\end{table}

\clearpage

\begin{table}[!t]
\centering
\caption{Fitted values of the resistance components $R_{\Omega}$ and $R_{ct}$, along with the calculated total conductivity ($\sigma_{\textrm{total}}$), for the NASICON compounds under test.}
\label{Rvalue}
\setlength{\tabcolsep}{6.1mm}{}
\renewcommand{\arraystretch}{1.1}
\begin{tabular}{llll}
\hline
Doped NASICONs                                                           & $R_{\Omega}$           &  $R_{ct}$           & $\sigma_{\textrm{total}}$         \\
                                                                         & $(\Omega)$             &  $(\Omega)$         & (S/cm)                            \\
\hline
Na$_{3.36}$Zr$_{1.92}$Cu$_{0.08}$Si$_{2.2}$P$_{0.8}$O$_{12}$             & 30.31                  &  35.35              &  $9.62 \times 10^{-4}$            \\
Na$_{3.1}$Zr$_{1.9}$Nd$_{0.1}$Si$_{2}$PO$_{12}$                          & 49.32                  &  53.21              &  $5.48 \times 10^{-4}$            \\
Na$_{3.2}$Zr$_{1.8}$Fe$_{0.2}$Si$_{2}$PO$_{12}$                          & 66.99                  &  263.8              &  $1.68 \times 10^{-4}$            \\
\hline
\end{tabular}
\end{table}

\end{document}